# Radial Deformation Emplacement in Power Transformers Using Long Short-Term Memory Networks

Arash Moradzadeh
*Faculty of Electrical and Computer Engineering,*
*University of Tabriz,*
Tabriz, Iran
arash.moradzadeh@ieee.org

Kazem Pourhossein
*Department of Electrical Engineering,*
*Tabriz Branch,*
*Islamic Azad University,*
Tabriz, Iran
k.pourhossein@iaut.ac.ir

Behnam Mohammadi-Ivatloo
*Faculty of Electrical and Computer Engineering,*
*University of Tabriz,*
Tabriz, Iran
mohammadi@ieee.org

Tohid Khalili
*Department of Electrical and Computer Engineering,*
*University of New Mexico,*
Albuquerque, USA
khalili@unm.edu

Ali Bidram
*Department of Electrical and Computer Engineering,*
*University of New Mexico,*
Albuquerque, USA
bidram@unm.edu

*Abstract*— A power transformer winding is usually subject to mechanical stress and tension because of improper transportation or operation. Radial deformation (RD) is an example of mechanical stress that can impact power transformer operation through short circuit faults and insulation damages. Frequency response analysis (FRA) is a well-known method to diagnose mechanical defects in transformers. Despite the precision of FRA, the interpretation of the calculated frequency response curves is not straightforward and requires complex calculations. In this paper, a deep learning algorithm called long short-term memory (LSTM) is used as a feature extraction technique to locate RD faults in their early stages. The experimental results verify the effectiveness of the proposed method in the diagnosis and locating of RD defects.

*Keywords*— *Deep learning, frequency response analysis, long short-term memory, radial deformation, transformer winding.*

## I. INTRODUCTION

Power transformers are critical and expensive equipment in power systems that play a vital role in the reliable delivery of power [1]. Power transformers are exposed to various mechanical defects due to improper transportation, fault conditions, inrush currents, the incidence of impulse voltages, etc [2]-[3]. Timely detection of these faults and mechanical defects is of paramount value to ensure the reliable operation of the power system. According to [4], on-line tap changer (OLTC) and winding faults have the highest probability compared to other fault types (See Fig. 1). Compared to the other parts of the transformer, identifying and repairing winding faults is more a challenging and time-consuming task. The mechanical defects in windings can result in insulation damage which in turn lead to short circuit (SC) faults [5]-[12]. Radial deformation (RD) of winding is one of the common defects in transformers. RD can happen due to the radial mechanical stresses imposed by winding. Fig. 2 shows the impact of RD fault on the transformer winding. RD fault can significantly change the distribution of ground capacitance while the other parameters such as series capacitances and inductances stay intact [13].

Various methods such as winding impedance measurement [14], transformer vibration analysis [15], dissolved gas analysis (DGA) [16], and frequency response analysis (FRA) [17]–[19] have been proposed to diagnose the mechanical faults and changes in transformer windings. According to [20]-[21], FRA offers a more accurate diagnosis compared to the other aforementioned methods. FRA relies on the fact that any mechanical anomaly in the active section of a transformer leads to the variation of its electric parameters which in turn changes the frequency response signatures.

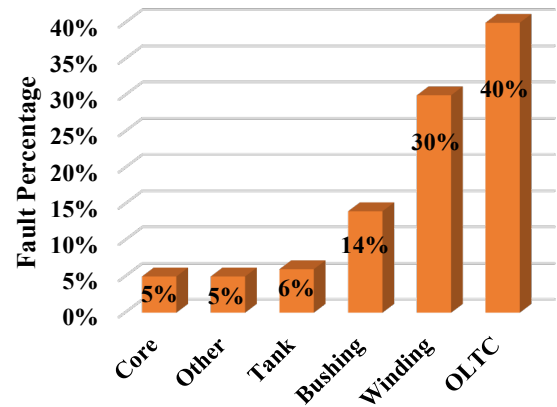

Fig. 1. Power transformer faults' probability percentage.

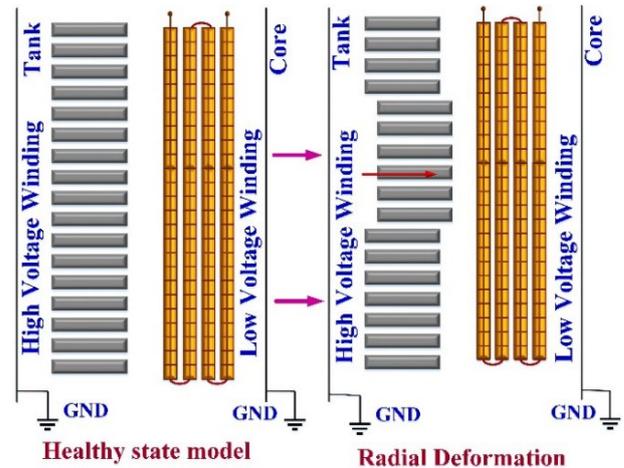

Fig. 2. Impact of RD fault on the winding.

However, interpreting the frequency response signatures is a challenging task and requires complex data analytic methods.

Tohid Khalili and Ali Bidram are supported by the National Science Foundation EPSCoR Program under Award #OIA-1757207.

To date, several solutions have been proposed to interpret the frequency response of the power transformer for diagnosing the winding defects. The SC fault detection in transformer windings has been performed using finite element analysis of frequency responses in [8]. Interpretation of frequency responses calculated from the SC faults in transformer windings is presented in [12] using statistical indices. In [9], a traditional neural network called Multilayer Perceptron (MLP) has been employed to interpret the frequency responses associated with short-circuit faults in transformer windings. In another valuable work [11], one of the machine learning application called support vector regression (SVR) interprets the frequency responses of SC faults to estimate the location of faults. In [5] and [6], the statistical and mathematical indices have been selected for calculating the salient deviations from frequency response in order to identify the type and location of mechanical faults in transformer windings.

In [22], numerical indices are used to identify mechanical defects. A vector-based method using principal component analysis (PCA) is used in [23] to interpret frequency response curves and discriminate mechanical defects in transformer windings. In [24], FRA polar plot along with digital image processing techniques are used to interpret frequency response signatures. In [25], one of the manifold learning application called Isometric feature mapping (Isomap) has been selected to interpret the frequency responses and locating inter-turn SC faults in transformer windings. In [26], the parameters of the transformer model have been estimated for mechanical fault localization. Support vector machine (SVM) is used in [7] and [27] for transfer function analysis and classifying fault types. In [28], the cross-correlation method is selected for the interpretation of estimated frequency responses. Enhanced magnetic optimal procedure to estimate fault location from sweep frequency response analysis is proposed in [29]. Assessment of binary image obtained from frequency response signatures to detect mechanical defects is performed in [30]. Improved FRA-based binary morphology and extreme point variation are used in [31]. In [32], the finite element method is employed to interpret frequency response curves for fault diagnosis. It should be noted that in the aforementioned frequency response analysis methods, the fault detection has been done after the fault has completely occurred and damaged the transformers. Moreover, some of these methods fail to identify the exact location of the fault.

In this paper, a deep learning algorithm, namely long short-term memory (LSTM), is used to extract the features of winding frequency response for the early detection of RD faults in transformer windings. The high performance of the LSTM in processing high-dimensional data such as frequency response curves obtained from transformer windings is one of the most important reasons for choosing this method. Since traditional neural networks suffer from over-fitting problems in the face of the high-dimensional data, the LSTM technique overcomes this problem due to its time-series structure and becomes a suitable replacement for traditional neural networks. LSTM is applied to frequency response curves of the transformer winding to detect the exact location of RD in the experimental setup. To this end, a training set based on the frequency response curves is formed to train LSTM. The trained network is capable of detecting low-intensity RD faults.

The remainder of this paper is organized as follows: Section II describes FRA. Section III introduces the LSTM. In section IV, the experimental setup is presented. Experimental results are presents in section V. Finally, Section VI concludes the paper.

## II. FREQUENCY RESPONSE ANALYSIS

FRA as a comparison-based method is an efficient and sensitive tool to evaluate the mechanical integrity of power transformers. To detect the abnormalities of the winding, the frequency response curve of the intact state of winding is required [18]-[19]. The principle idea of FRA is to assess the effect of any mechanical variation of transformer that affects the frequency response signatures. The common types of FRA are low-voltage impedance (LVI) or sweep frequency response analysis (SFRA) methods [19]-[20]. In this paper, the LVI method is used to obtain frequency responses. In this method, the input and output signals are calculated by applying a surge voltage to the transformer and then measuring the current. The Frequency Response (FR) is defined as [33]

$$FR = \left| \frac{I_o(f)}{V_i(f)} \right| \quad (1)$$

where $I_o(f)$ is the ground current measured at the end of the winding and $V_i(f)$ is the input voltage in the frequency space.

## III. LONG SHORT-TERM MEMORY

LSTM is a standard and efficient recurrent neural network (RNN) architecture proposed by Hochreiter & Schmidhuber for modeling and forecasting time series [34]-[35]. Nowadays, LSTM is used as a powerful tool in deep learning applications to improve RNN structures, regression, and classification. Traditional neural networks are trained in such a way that they only learn the relationship between input variables and targets from a static viewpoint. Compared to traditional neural networks, RNNs communicate between both input and output pairs [36]-[37]. Problems of the gradient disappearance and gradient explosion in the training of traditional RNN are solved by the gate mechanism and memory unit of LSTM. As shown in Fig. 3, the forget $f_{(t)}$, input $i_{(t)}$, update $g_{(t)}$, and output $o_{(t)}$ gates form the LSTM unit structure. Each of these gates has a unique function. Given the input vectors as $X = (x_1, x_2, ..., x_{t-1}, x_t)$ and output vectors as $Y = (y_1, y_2, ..., y_{t-1}, y_t)$, the activation values for gates and memory cell are calculated continuously with respect to the memory time t by the recursive hidden layer. The LSTM forget gate formula can be expressed as [38]-[39]:

$$f_t = \sigma( w_f * [h_{t-1}, x_t] + b_f ) \quad (2)$$

where $f_t$ is the output value of the forget gate at time t; $\sigma$ shows the sigmoid activation function; $w_f$ is the weight matrix of the forget gate; * denotes the appending operator for vectors; $h_{t-1}$ is the output vector at time t – 1; $b_f$ is the bias of the forget gate at time t. In the next gate, which is the input gate, the current information that should be used as an input to generate the memory cell $c_t$ is calculated as [36], [37], [39]

$$f_i = \sigma( w_i * [h_{t-1}, x_t] + b_i ) \quad (3)$$

where $w_i$ is the weight matrix of the input gate and $b_i$ denotes the bias in this gate. After determining the input gate, $c_t$ is formed using:

$$g_t = tanh(w_g * [h_{t-1}, x_t] + b_g) \quad (4)$$

$$c_t = f_t * c_{t-1} + i_t * g_t \quad (5)$$

where $g_t$ is the update gate; $w_g$ and $b_g$ denote the weight matrix and update gate bias, respectively. $f_t$ controls the long-term information, and it controls the short-term information. In the next step, the output gate $o_t$ is calculated as [34], [36]

$$o_t = \sigma(w_o * [h_{t-1}, x_t] + b_o) \quad (6)$$

where $w_o$ and $b_o$ are the weight matrix and bias of the output gate. Finally, the final output of LSTM is defined as [34]

$$h_t = o_t * tanh(c_t) \quad (7)$$

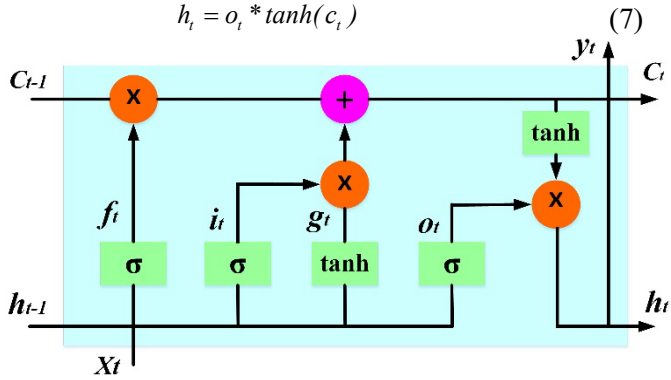

Fig. 3. Principle structure of an LSTM unit

After calculating the outputs and locating RD faults via LSTM, the results are evaluated using statistical performance metrics such as correlation coefficient (R), mean squared error (MSE), and root mean squared error (RMSE). Each of the metrics indicators for $T$ number of inputs is calculated using the following [37]

$$R = \frac{\sum_{t=1}^{T}(X_t - \bar{X})(Y_t - \bar{Y})}{\sqrt{\sum_{t=1}^{T}(X_t - \bar{X})^2 \sum_{t=1}^{N}(Y_t - \bar{Y})^2}} \quad (8)$$

$$MSE = \frac{1}{T}\sum_{t=1}^{T}(X_t - Y_t)^2 \quad (9)$$

$$RMSE = \sqrt{\frac{1}{T}\sum_{t=1}^{T}(X_t - Y_t)^2} \quad (10)$$

where $X_t$, $\bar{X}$, $Y_t$, and $\bar{Y}$ denote the real value, mean of real values, predicted value, and average of real values, respectively.

IV. EXPERIMENTAL SETUP

In this paper, we studied a single-phase transformer. All the tests were carried out on the high voltage (HV) winding, and the low voltage winding (LV) was in open circuit state. The HV winding consists of 160 rounds of wire in diameter of 7 mm around the core. The length of this winding is 120 cm (See Fig. 4). Every 40 rounds of HV winding is considered as a section which results in four sections. Three different RD faults were applied to each section with intensities of 6, 9, and 12 mm per section. Fig. 5 shows how an RD fault was applied to the winding. In this experiment, the GPS-1102B oscilloscope was used to measure and save the signals.

V. EXPERIMENTAL RESULTS

In this section, first, the frequency response related to the healthy state of the transformer is calculated using (1). This will be later compared with the frequency responses of fault conditions. Then, the RD faults explained in Section IV are applied and the frequency response corresponding to each of them is calculated. Fig. 6 presents the frequency response curves for healthy state and the radial deformation faults with 3 intensities in section 3 of the winding. Also, Fig. 7 shows the frequency response curves for the healthy state and RD fault by the intensity of 12 mm in all 4 sections of the winding. LSTM is used to interpret the frequency responses and extract their unique features. Applying the LSTM method requires a dataset as input. Frequency responses related to all RD faults were collected as a dataset and considered as inputs. The location of faults was considered as a target regardless of its severity. For the training phase of the network, frequency responses for faults with the intensity of 9 and 12 mm were used. So, 8 samples of frequency responses (4 × 2), each of which with a frequency range of 2.5 MHz, formed the input dataset. The network was trained and validated in one step with the input dataset. Then, the frequency responses for faults with an intensity of 6 mm were used for the final test of the LSTM. Fig. 8 shows the results of fault detection according to the target set for each fault via the LSTM network in the training phase. Fig 9 shows the results of fault detection in the initial testing phase of the trained network using data randomly assigned by the network itself. Figs. 8 and 9 show the accuracy of the designed network after training and test phases and indicate that the designed network can identify minor RD defects with excellent correlation (R=1) and zero MSE and RMSE.

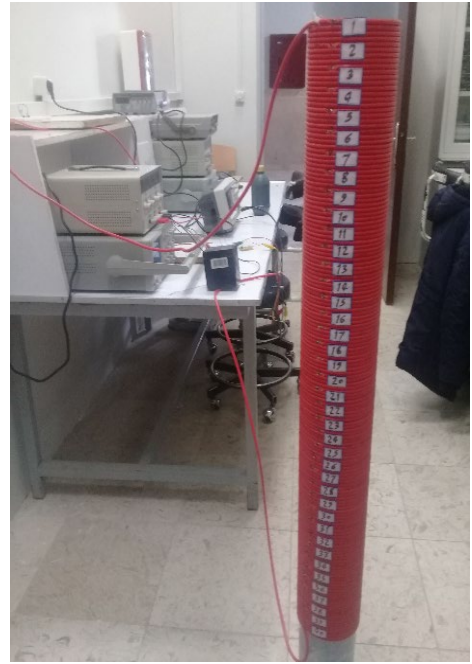

Fig. 4. Experimental setup.

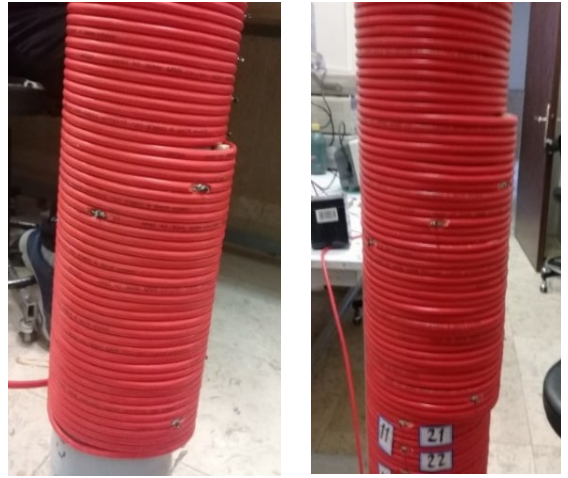

Fig. 5. RD fault implementation on the winding.

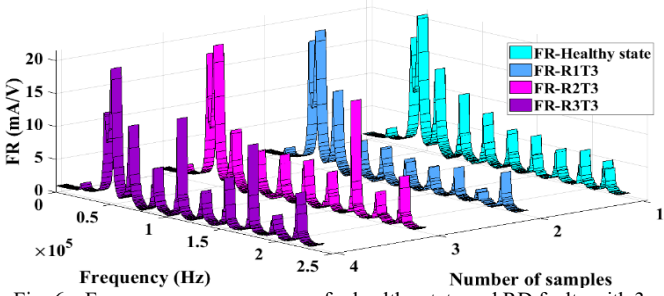

Fig. 6. Frequency response curves for healthy state and RD faults with 3 intensities in section 3

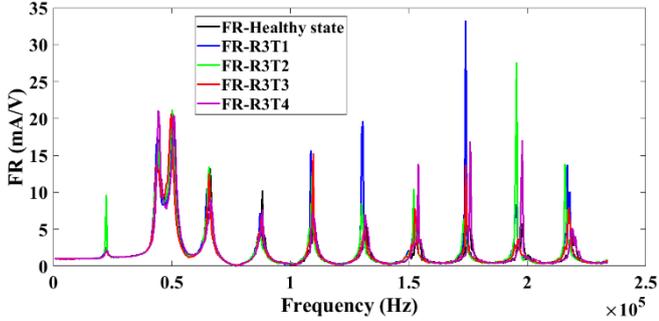

Fig. 7. Frequency response curves for healthy state and RD fault by the intensity of 12 mm in all sections

The trained LSTM acts as a fault diagnosis toolbox containing features extracted from frequency responses related to RD faults. To verify the performance of this toolbox, the frequency responses of RD faults with 6 mm intensity are used. Since the frequency responses associated with the 6 mm RD faults were not included in the LSTM input dataset, they can be considered as new faults for test the trained network. Fig. 10 shows the results of the designed LSTM prediction in identifying the location of new RD faults (early-stages). The presented results verify the effectiveness of the designed LSTM network in identifying and locating low-intensity RD faults (6 mm severity) with excellent correlation (R=1) and zero MSE and RMSE.

As mentioned in the literature, the method suggested in this paper is applied to data related to a single-phase of a three-phase transformer winding. Similarly, the proposed procedure can be implemented to identify faults corresponding to each phase of the transformer. To do this, it is enough to calculate the frequency response related to the fault generated in each phase and consider it as the input of the LSTM technique. Fault detection after the complete occurrence and lack of accurate diagnosis can, in addition to serious damage to the transformer itself, reduce the reliability of the power system, and cause unwanted blackouts. Accordingly, the proposed LSTM technique in this paper with its high capability in early detection of faults will be able to prevent the occurrence of the mentioned damages in the power system. Also, the LSTM can be used as a tool for online monitoring applications in power systems.

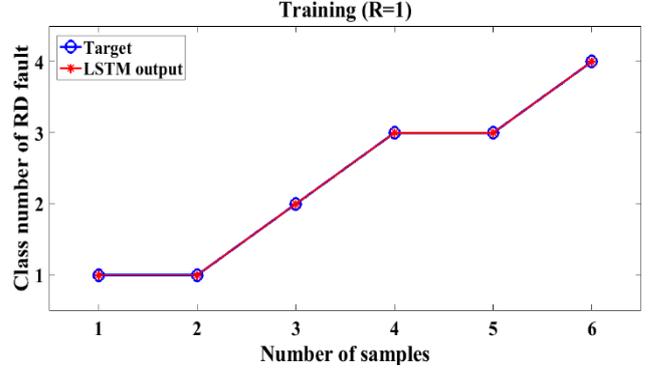

Fig. 8. Classification of RD faults by LSTM using train data.

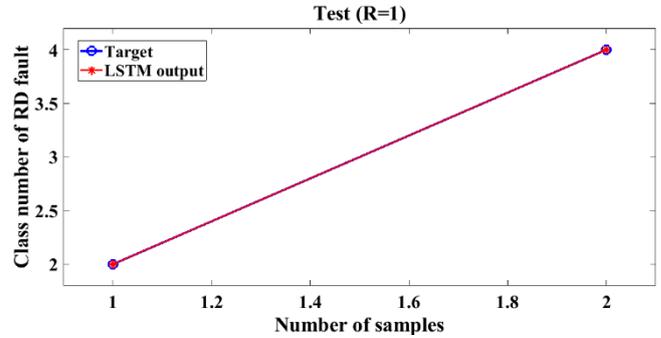

Fig. 9. Classification of RD faults by trained LSTM using test data.

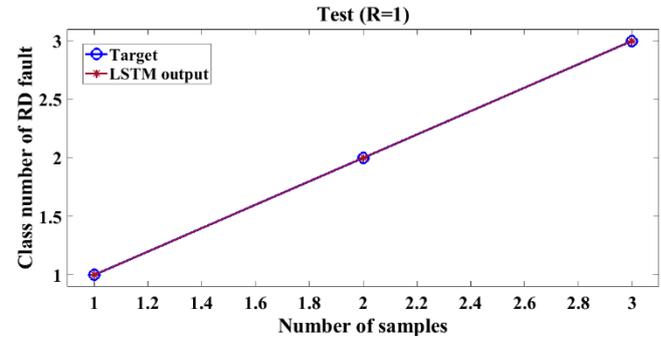

Fig. 10. Classification of RD faults by trained LSTM using unknown data (new faults).

## VI. CONCLUSION

FRA is one of the methods for detecting mechanical faults on transformer windings. However, the interpretation of frequency response curves is a challenging and time-consuming task. To tackle this challenge, this paper utilizes LSTM to evaluate and interpret frequency response curves of power transformers. LSTM is applied to frequency response curves of transformer windings to detect the exact location of RD. The trained LSTM is used to detect new RD faults with different intensities. An experimental setup is used to verify the effectiveness of the proposed method. The results show the high accuracy of LSTM in identifying transformer winding RD faults.

# REFERENCES

[1] S. A. Saleh *et al.*, "Solid-state transformers for distribution systems-Part II: Deployment challenges," *IEEE Trans. Industry Applications*, vol. 55, no. 6, pp. 5708–5716, 2019.

[2] Z. Liang and A. Parlikad, "A markovian model for power transformer maintenance," *International Journal of Electrical Power and Energy Systems*, vol. 99, pp. 175–182, 2018.

[3] W. S. Fonseca, D. S. Lima, A. K. F. Lima, M. V. A. Nunes, U. H. Bezerra, and N. S. Soeiro, "Analysis of structural behavior of transformer's winding under inrush current conditions," *IEEE Trans. Industry Applications*, vol. 54, no. 3, pp. 2285–2294, 2018.

[4] C. AJ, M. A. Salam, Q. M. Rahman, F. Wen, S. P. Ang, and W. Voon, "Causes of transformer failures and diagnostic methods – A review," *Renewable and Sustainable Energy Reviews*, vol. 82. pp. 1442–1456, 2018.

[5] E. Rahimpour, M. Jabbari, and S. Tenbohlen, "Mathematical comparison methods to assess transfer functions of transformers to detect different types of mechanical faults," *IEEE Trans. on Power Delivery*, vol. 25, no. 4, pp. 2544–2555, 2010.

[6] E. Rahimpour, J. Christian, K. Feser, and H. Mohseni, "Transfer function method to diagnose axial displacement and radial deformation of transformer windings," *IEEE Trans. on Power Delivery*, vol. 18, no. 2, pp. 493–505, 2003.

[7] H. Tarimoradi and G. B. Gharehpetian, "Novel calculation method of indices to improve classification of transformer winding fault type, location, and extent," *IEEE Trans. on Industrial Informatics*, vol. 13, no. 4, pp. 1531–1540, 2017.

[8] H. M. Ahn, J. Y. Lee, J. K. Kim, Y. H. Oh, S. Y. Jung, and S. C. Hahn, "Finite-element analysis of short-circuit electromagnetic force in power transformer," *IEEE Trans. on Industry Applications*, vol. 47, no. 3, pp. 1267–1272, 2011.

[9] A. Moradzadeh and K. Pourhossein, "Early detection of turn-to-turn faults in power transformer winding: An experimental study," 2020, pp. 199–204.

[10] E. Rahimpour and S. Tenbohlen, "Experimental and theoretical investigation of disc space variation in real high-voltage windings using transfer function method," *IET Electric Power Applications*, vol. 4, no. 6, pp. 451–461, 2010.

[11] A. Moradzadeh and K. Pourhossein, "Application of support vector machines to locate minor short circuits in transformer windings," in *2019 54th International Universities Power Engineering Conference, UPEC 2019 - Proceedings*, 2019, pp. 1–6.

[12] J. N. Ahour, S. Seyedtabaii, and G. B. Gharehpetian, "Determination and localisation of turn-to-turn fault in transformer winding using frequency response analysis," *IET Science, Measurement and Technology*, vol. 12, no. 3, pp. 291–300, 2018.

[13] P. Mukherjee and L. Satish, "Localization of radial displacement in an actual isolated transformer winding-An analytical approach," *IEEE Trans. on Power Delivery*, vol. 31, no. 6, pp. 2511–2519, 2016.

[14] T. Chiulan and B. Pantelimon, "A practical example of power transformer unit winding condition assessment by means of short-circuit impedance measurement," in *2009 IEEE Bucharest PowerTech: Innovative Ideas Toward the Electrical Grid of the Future*, 2009.

[15] H. Zhou, K. Hong, H. Huang, and J. Zhou, "Transformer winding fault detection by vibration analysis methods," *Applied Acoustics*, vol. 114, pp. 136–146, 2016.

[16] H. Malik and S. Mishra, "Application of Gene Expression Programming (GEP) in Power Transformers Fault Diagnosis Using DGA," *IEEE Trans. on Industry Applications*, vol. 52, no. 6, pp. 4556–4565, 2016.

[17] A. Moradzadeh and K. Pourhossein, "Location of Disk Space Variations in Transformer Winding using Convolutional Neural Networks," in *2019 54th International Universities Power Engineering Conference, UPEC 2019 - Proceedings*, 2019.

[18] R. Khalili Senobari, J. Sadeh, and H. Borsi, "Frequency response analysis (FRA) of transformers as a tool for fault detection and location: A review," *Electric Power Systems Research*, vol. 155. pp. 172–183, 2018.

[19] S. A. Ryder, "Transformer diagnosis using frequency response analysis: Results from fault simulations," in *Proceedings of the IEEE Power Engineering Society Transmission and Distribution Conference*, 2002, vol. 1, no. SUMMER, pp. 399–404.

[20] J. C. Gonzales and E. E. Mombello, "Fault interpretation algorithm using frequency-response analysis of power transformers," *IEEE Trans. on Power Delivery*, vol. 31, no. 3, pp. 1034–1042, 2016.

[21] K. Pourhossein, G. B. Gharehpetian, E. Rahimpour, and B. N. Araabi, "A probabilistic feature to determine type and extent of winding mechanical defects in power transformers," *Electric Power Systems Research*, vol. 82, no. 1, pp. 1–10, 2012.

[22] K. Pourhossein, G. B. Gharehpetian, and E. Rahimpour, "Buckling severity diagnosis in power transformer windings using Euclidean Distance classifier," in *2011 19th Iranian Conference on Electrical Engineering, ICEE 2011*, 2011.

[23] K. Pourhossein, G. B. Gharehpetian, E. Rahimpour, and B. N. Araabi, "A vector-based approach to discriminate radial deformation and axial displacement of transformer winding and determine defect extent," *Electric Power Components and Systems*, vol. 40, no. 6, pp. 597–612, 2012.

[24] A. Abu-Siada and O. Aljohani, "Detecting incipient radial deformations of power transformer windings using polar plot and digital image processing," *IET Science, Measurement and Technology*, vol. 12, no. 4, pp. 492–499, 2018.

[25] A. Moradzadeh, K. Pourhossein, B. Mohammadi-Ivatloo, and F. Mohammadi, "Locating Inter-Turn Faults in Transformer Windings Using Isometric Feature Mapping of Frequency Response Traces," IEEE Transactions on Industrial Informatics, pp. 1–1, 2020, doi: 10.1109/TII.2020.3016966.

[26] L. Satish and S. K. Sahoo, "Locating faults in a transformer winding: An experimental study," *Electric Power Systems Research*, vol. 79, no. 1, pp. 89–97, 2009.

[27] M. Bigdeli, M. Vakilian, and E. Rahimpour, "Transformer winding faults classification based on transfer function analysis by support vector machine," *IET Electric Power Applications*, vol. 6, no. 5, pp. 268–276, 2012.

[28] A. R. Abbasi, M. R. Mahmoudi, and Z. Avazzadeh, "Diagnosis and clustering of power transformer winding fault types by crosscorrelation and clustering analysis of FRA results," *IET Generation, Transmission and Distribution*, vol. 12, no. 19, pp. 4301–4309, 2018.

[29] M. S. Jahan, R. Keypour, H. R. Izadfar, and M. T. Keshavarzi, "Locating power transformer fault based on sweep frequency response measurement by a novel multistage approach," *IET Science, Measurement and Technology*, vol. 12, no. 8, pp. 949–957, 2018.

[30] Z. Zhao, C. Yao, C. Tang, C. Li, F. Yan, and S. Islam, "Diagnosing transformer winding deformation faults based on the analysis of binary image obtained from FRA signature," *IEEE Access*, vol. 7, pp. 40463–40474, 2019.

[31] Z. Zhao, C. Yao, C. Li, and S. Islam, "Detection of power transformer winding deformation using improved FRA based on binary morphology and extreme point variation," *IEEE Trans. on Industrial Electronics*, vol. 65, no. 4, pp. 3509–3519, 2018.

[32] J. Jiang, L. Zhou, S. Gao, W. Li, and D. Wang, "Frequency response features of axial displacement winding faults in autotransformers with split windings," *IEEE Trans. on Power Delivery*, vol. 33, no. 4, pp. 1699–1706, 2018.

[33] A. Moradzadeh and K. Pourhossein, "Short circuit location in transformer winding using deep learning of its frequency responses," 2020, pp. 268–273.

[34] Y. Wang, D. Gan, M. Sun, N. Zhang, Z. Lu, and C. Kang, "Probabilistic individual load forecasting using pinball loss guided LSTM," *Applied Energy*, vol. 235, pp. 10–20, 2019.

[35] S. Hochreiter and J. Schmidhuber, "Long short-term memory," *Neural Computation*, vol. 9, no. 8, pp. 1735–1780, 1997.

[36] A. Moradzadeh, S. Zakeri, M. Shoaran, B. Mohammadi-Ivatloo, and F. Mohamamdi, "Short-term load forecasting of microgrid via hybrid support vector regression and long short-term memory algorithms," Sustainability (Switzerland), vol. 12, no. 17, p. 7076, Aug. 2020, doi: 10.3390/su12177076.

[37] Y. Wu, Q. Xue, J. Shen, Z. Lei, Z. Chen, and Y. Liu, "State of health estimation for lithium-Ion batteries based on healthy features and long short-term memory," *IEEE Access*, vol. 8, pp. 28533–28547, 2020.

[38] X. Yuan, L. Li, and Y. Wang, "Nonlinear dynamic soft sensor modeling with supervised long short-term memory network," *IEEE Trans. on Industrial Informatics*, vol. 16, no. 5, pp. 3168–3176, 2019.

[39] L. Han, H. Jing, R. Zhang, and Z. Gao, "Wind power forecast based on improved Long Short Term Memory network," *Energy*, vol. 189, 2019.